\documentclass[rnote]{aa} 
\usepackage{amssymb,graphicx,graphics}

\begin{document}
\title{Deep search for companions to probable young brown dwarfs\thanks{Based on observations collected at the European Southern Observatory, Chile (ESO programmes 076.C-0554(A) 
, 076.C-0554(B) and 085.C-0257(A)}}
\subtitle{VLT/NACO adaptive optics imaging using IR wavefront sensing}
\author{G. Chauvin\inst{1,2}\and
        J. Faherty\inst{3,4}\and
        A. Boccaletti\inst{5}\and
        K. Cruz\inst{4,6}\and
        A.-M. Lagrange\inst{1}\and
        B. Zuckerman\inst{7}\and
	M.\,S. Bessell\inst{8}\and
        J.-L. Beuzit\inst{1}\and
        M. Bonnefoy\inst{2}\and
        C. Dumas\inst{9}\and
        P. Lowrance\inst{10}\and
        D. Mouillet\inst{1}\and
	I. Song\inst{11}
}
\institute{
$^{1}$UJF-Grenoble1/CNRS-INSU, Institut de Plan\'etologie et d'Astrophysique de Grenoble UMR 5274, Grenoble, F-38041, France\\
$^{2}$Max Planck Institute for Astronomy, K\"onigstuhl 17, D-69117 Heidelberg, Germany\\
$^{3}$Departmento de Astronomia, Universidad de Chile, Casilla 36-D, Santiago, Chile\\
$^{4}$Department  of Astrophysics, American Museum of Natural History, Central Park West at 79th Street, New York, NY, 10024\\
$^{5}$LESIA-Observatoire de Paris, CNRS, UPMC Univ. Paris 06, Univ. Paris-Diderot, 92195, Meudon, France\\
$^{6}$Department of Physics \& Astronomy, Hunter College, City University of New York, 695 Park Avenue, New York, NY 10065, USA\\
$^{7}$Department of Physics \& Astronomy and Center for Astrobiology, University of California: Los Angeles, Box 951562, CA 90095, USA\\
$^{8}$Research School of Astronomy and Astrophysics Institute of Advance Studies, Australian National University: Cotter Road, Weston Creek, Canberra, ACT 2611, Australia\\
$^{9}$European Southern Observatory: Casilla 19001, Santiago 19, Chile\\
$^{10}$ Infrared Processing and Analysis Center, MS 100-22, California Institute of Technology, Pasadena, CA 91125
$^{11}$Department of Physics \& Astronomy, University of Georgia, Athens, GA 30602-2451, USA
}

\offprints{G. Chauvin}
\date{Received ; accepted }

  \abstract 
  {}
  {We have obtained high contrast images of four nearby, faint, and
    very low mass objects 2MASS\,J04351455-1414468,
    SDSS\,J044337.61+000205.1, 2MASS\,J06085283-2753583 and
    2MASS\,J06524851-5741376 (here after 2MASS0435-14, SDSS0443+00,
    2MASS0608-27 and 2MASS0652-57), identified in the field as
    probable isolated young brown dwarfs.  Our goal was to search for
    binary companions down to the planetary mass regime.}
  {We used the NAOS-CONICA adaptive optics instrument (NACO) and its
    unique capability to sense the wavefront in the near-infrared to
    acquire sharp images of the four systems in $K_s$, with a field of
    view of $28~\!''\times28~\!''$. Additional $J$ and $L'$ imaging
    and follow-up observations at a second epoch were obtained for 2MASS0652-57.}
   {With a typical contrast $\Delta K_s=4.0-7.0$~mag, our observations
     are sensitive down to the planetary mass regime considering a
     minimum age of 10 to 120~Myr for these systems. No additional
     point sources are detected in the environment of 2MASS0435-14,
     SDSS0443+00 and 2MASS0608-27 between $0.1-12~\!''$ (i.e about 2
     to 250~AU at 20~pc). 2MASS0652-57 is resolved as a $\sim230$~mas
     binary. Follow-up observations reject a background contaminate,
     resolve the orbital motion of the pair, and confirm with high
     confidence that the system is physically bound. The $J$, $K_s$
     and $L'$ photometry suggest a $q\sim0.7-0.8$ mass ratio binary
     with a probable semi-major axis of 5-6~AU. Among the four
     systems, 2MASS0652-57 is probably the less constrained in terms
     of age determination. Further analysis would be necessary to
     confirm its youth. It would then be interesting to determine its
     orbital and physical properties to derive the system's dynamical
     mass and to test evolutionary model predictions.}
   {}

   \maketitle

\section{Introduction}


The statistical properties of low-mass star and brown dwarf multiples
set stringent constraints on star-formation theories (see Duchene et
al. 2007 for a review).  Multiplicity frequency, mass ratio and
separation distributions can be compared between star forming regions
(SFR's) of various ages and densities and the older field
population. Direct imaging surveys of very low mass objects in the
field yield a binary frequency of 20-30\% for M dwarfs (Marchal et al. 2003; Janson et
al. 2012) and 15\% for L and T dwarfs (e.g., Bouy et al. 2003;
Burgasser et al. 2003). Among young systems, Ahmic et al. (2007) and
Biller et al. (2011) derive a binary fraction of less than 11\% and
9\% in the Chamaeleon I and the Upper Sco regions respectively. They
both confirm the trend observed in the field of a mass-dependency of
the binary frequency. A higher multiplicity rate in SFR's compared with the field, as for T Tauri stars,
is not seen for young late-type M dwarfs.  However the multiplicity
properties are likely to be different as evidenced by the discovery of
a population of wide ($>15$~AU) brown dwarf binaries in young, nearby
clusters (Chauvin et al. 2004; Jayawardhana \& Ivanov 2006; B\'ejar et
al. 2008; Todorov et al. 2010). Individual, young, and tight binaries
are particularly important for a direct determination of the dynamical
mass to calibrate theoretical masses derived from evolutionary models
(Mathieu et al. 2007; Bonnefoy et al. 2009).
\begin{table*}[t]
\caption{Description of the target properties}
\begin{center}
\small
\begin{tabular}{lllllllllll}     
\noalign{\smallskip}\hline
\noalign{\smallskip}\hline  \noalign{\smallskip}
Name            & $\alpha$        &  $\delta$       &    $\mu_{\alpha}^a$   & $\mu_{\delta}^a$     & Vrad       & SpT      & $d^a$     &   J     & K        & Ref.$^b$ \\
                & [J2000]         & [J2000]         &    (mas/yr)         & (mas/yr)       & (km/s)     &          & (pc)  &   (mag) & (mag)    &            \\
\noalign{\smallskip}\hline                  \noalign{\smallskip}
2MASS0435-14	& 04 35 14.6      & -14 14 47       & $9\pm14$            & $16\pm14$      &            &   M6$\delta\pm$1     & $8.6\pm1.0$       & 11.88  & 9.95    & 1 \\
SDSS0443+00	& 04 43 37.6      & +00 02 05       & $28\pm14$ 	  & $-99\pm14$     &            & M9$\gamma$      & $16.2\pm2.1$      & 12.51  & 11.22   & 1, 2, 3, 4\\
2MASS0608-27	& 06 08 52.8      & -27 53 58       & $8.9\pm3.5$         & $10.7\pm3.5$   & $24\pm1$   & M8.5$\gamma$     & $31.2_{-3.2}^{+4.0}$ & 13.59  & 12.37   & 1, 5, 6 \\
2MASS0652-57	& 06 52 48.5      & -57 41 38       & $0.1\pm3.4$ 	  & $29.2\pm3.3$   &            & M8$\beta$       & $31.9_{-2.9}^{+3.7}$ & 13.63  & 12.45   & 6, 7\\   
\noalign{\smallskip}\hline  \noalign{\smallskip}
\end{tabular}
\begin{list}{}{}
\item[$^{\mathrm{a}}$] Distances derived from spectrophotometry with M$_J$ estimated from the spectral type/M$_J$ calibration from Cruz et al. (2003) for 2MASS0435-14 and SDSS0443+00. Proper motion and parallax measurements from Faherty et al. (2012) for 2MASS0608-27 and 2MASS0652-57
\item[$^{\mathrm{b}}$] References: (1) Cruz et al. 2003, AJ, 126, 2421, (2) Cruz et al. 2007, AJ, 133, 43, (3) Reid et al. 2008, 136, 1290, (4) Reiner \& Basri 2009, AJ, 705, 1416, (5) Rice et al. 2010, ApJ, 715, 165, (6) Faherty et al. (2012) and (7) Reid et al. (2008)
\end{list}
\end{center}
\end{table*}

For very low mass binaries with high mass ratio, the mass of the
binary companion can enter the planetary mass regime. The origin of
such a population of planetary mass companions (PMCs) can be hard to
infer as the stellar and planetary formation mechanisms probably
overlap. From radial velocity surveys, the observed frequency of giant
planets around M dwarfs at small separation ($<3$~AU) is relatively
small ($f < 1-2~\%$; Bonfils et al. 2011) compared with solar-type
stars ($f < 6-9~\%$; Udry \& Santos 2007) and could indicate a mass
dependency of the core accretion mechanism efficiency to form giant
planets. At larger separations, the situation is less clear. We may
expect that alternative mechanisms to core accretion such as cloud or
disk fragmentation may form a population of planetary mass companions
such as 2M1207\,b (Chauvin et al. 2004). Consequently, despite the
fact that most exoplanet imaging surveys are now biased towards young,
intermediate mass stars, the search for PMCs around low-mass stars and
brown dwarfs remains important to understand how planetary formation
evolves with the stellar mass and the distance to the star (Delorme et
al. 2012).

In the course of our deep imaging survey of 88 young, nearby stars
with NACO at VLT (Chauvin et al. 2010), an additional sub-sample of
four probable intermediate-young brown dwarfs were observed taking
advantage of the infrared (IR) wavefront sensing system of the NACO adaptive
optics (AO) instrument. One is a recently confirmed member of the
$\beta$ Pictoris moving group (Rice et al. 2010) while the other 3 are
low-gravity M dwarfs indicating a likely age lower than a few hundred
Myr. We report, in section 2, a summary of the target properties. In
section 3, we describe our observations, including the instrument
setup and the atmospheric conditions. In section 4, we present the
results of this imaging campaign, in terms of detection limits. We
also report that 2MASS0652-57 is a $\sim230$~mas binary. Observations at 2 epochs enable us to resolve the system
orbital motion. Finally, we briefly discuss the status of this young
binary, and of its membership to any known young moving groups.

\section{Target Properties}

\begin{figure}[t]
\centering
\vspace{-0.1cm}
\includegraphics[width = 6.5cm]{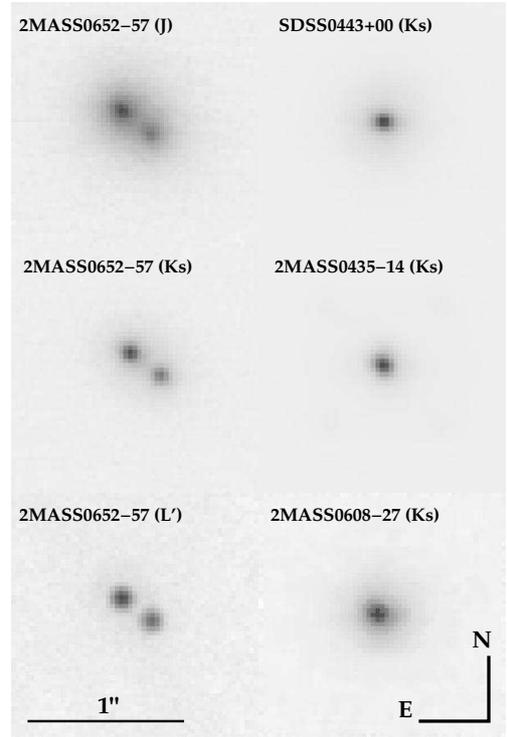}\hspace{0.5cm}
\caption{VLT/NACO image obtained of 2MASS0652-57 in $J$, $K_s$ and
  $L'$-bands and SDSS0443+00, 2MASS0608-27 and 2MASS0435-14 in $K_s$
  band. 2MASS0652-57 is resolved in each band as a close visual
  binary. All images were normalized to their maximum intensity.}
\label{fig:image}
\end{figure}
\begin{table*}[t]
\caption{Detail of the observing parameters. DIT and NDIT are the integration time and the number of integrations, respectively.}
\begin{center}
\small
\begin{tabular}{lllllllll}     
\noalign{\smallskip}\hline
\noalign{\smallskip}\hline  \noalign{\smallskip}
Name                  & UT Date               & Filter/obs            &      DIT      &      NDIT   & NINT     & Airmass  & Strehl   & \textit{FWHM}        \\
                      &                       &                       &      (s)      &             &          &          & (\%)     & (mas)       \\
\noalign{\smallskip}\hline                  \noalign{\smallskip}
SDSS0443+00           & 2006-08-01            & $K_s$/S27                &   30.         &  1          & 11       & 1.29     & 10      & 94          \\
                      & 2006-02-27            & $K_s$/S27                &   60.         &  1          & 3        & 1.27     & 5       & 130        \\
2MASS0652-57          & 2006-08-01            & $K_s$/S27                &   30.         &  2          & 5        & 1.28     & 9       & 95          \\
                      &                       & $J$/S27                 &   30.         &  2          & 5        & 1.31     & $<5$    & 200         \\
                      &                       & $L~\!'$/L27                &   0.175       & 100         & 21       & 1.25     & 15      & 118         \\
                      & 2010-08-21            & $K_s$/S27                &   30.         &  2          & 10       & 1.62     & 12      & 90          \\
2MASS0608-27	      & 2006-02-27            & $K_s$/S27                &   60.         &  1          & 10       & 1.03     & 5       & 130        \\ 
2MASS0435-14	      & 2006-02-27            & $K_s$/S27                &   30.         &  2          & 20       & 1.12     & 10      & 95         \\
\noalign{\smallskip}\hline  \noalign{\smallskip}
\end{tabular}
\label{tab:table2}
\end{center}
\end{table*}

2MASS0435-14, SDSS0443+00, 2MASS0652-57, and 2MASS0608-27 were
selected as imaging candidates because they have spectral or
photometric features indicative of youth (as described below).  To
date, only 2MASS0608-27 has been confirmed as a $\beta$ Pictoris
moving group member. Their properties are summarized in Table~1. As
suggested by Kirkpatrick (2005) and further discussed by Kirkpatrick
et al. (2006), we indicate spectra with low gravity indicators with
greek suffixes: subtle low-gravity features (Pleiades-like) are
indicated with a $\beta$, prominent low-gravity features (TWA-like)
are indicated with a $\gamma$, while very young (Taurus-like) objects
are indicated with a $\delta$. An $\alpha$ suffix is implied for
normal-gravity objects. We used these guidelines to assign spectral
types to these young/low-gravity M dwarfs even though the low-gravity
M dwarf spectral-typing scheme has yet to be formalized.

\begin{itemize}

\item \textit{2MASS0435-14}: This object was reported as a young
  object by Cruz et al. (2003; see Table~11 and Fig.~10) based on
  spectral features similar to young objects: weak CaH and K\,I
  doublet absorption and Na\,I and H$_\alpha$ emission. It is
  flagged as a photometric and kinematic outlier in Faherty et al
  (2009) due to its unusual red near-infrared color and very small tangential
  velocity (v$_{tan}=1\pm1$ km/s). The low-gravity spectral features
  of 2MASS0435-14 are similar to those seen in members of very young
  regions like Taurus. Visual inspection of the 2MASS and IRAS images
  reveals a significant amount of extinction in the vicinity of the
  object and it is near the MBM 20/LDN 1642 molecular cloud. Taking
  reddening into account, this object's red color is likely due to
  this material rather than condensates present in the object's
  atmosphere. The spectrum most resembles CHSM1982, an M6 object in
  Chameleon (Luhman 2004), and thus we tentatively
  assign a spectral type of M6$\delta\pm$1. This object warrants
  further investigation as it is likely a very young, very low mass
  object that appears to be isolated in a molecular cloud.
\item \textit{SDSS0443+00}: Intially classified as a field, old late-M
  dwarf by Cruz et al. (2003), this object was re-classified by Cruz
  et al. (2007; see Table~8 and Fig.~5) as it displays low gravity
  features similar to those seen in late-M type TWA members. Its
  optical spectrum is similar to Roque 4, an M9 Pleiad (Zapatero Osorio
  et al. 1998), but seems to have a lower gravity. We therefore assign a spectra type
  of M9$\gamma$. 
\item \textit{2MASS0608-27}: This object was first identified by Cruz
  et al. (2003) as a low-gravity and young brown dwarf candidate based
  on enhanced spectral VO absorption, as well as weak CaH and K\,I doublet
  absorption compared with older field dwarfs. More recently, a
  detailed study of its near-infrared spectral features (highly peaked $H$-band
  shape, intermediate widths and depth of the gravity-sensitive KI
  lines) and kinematics and space location, unambiguousy confirmed it as
  the latest-type non companion member of the
  $\beta$ Pictoris group. Considering an age of 12 Myr and an assigned spectral type of M8.5gamma (Rice et al. 2010), the estimated mass of 15-20 MJup.
\item \textit{2MASS0652-57}: For this system, suspicion of youth was
  based on a low-resolution optical spectrum that shows hints of
  low-gravity, including enhanced spectral VO absorption and weak
  Na\,I and CaH absorptions. The low-gravity features are similar to those
  seen in late-M type Pleiades members, and indeed its optical
  spectrum looks very similar to Roque 7, an M8-type Pleiad (Mart\'in et
  al. 2000). We assign a spectra type of M8$\beta$.
\end{itemize}


\section{Observations and data reduction}

On January 8th and February 27th 2006, the four brown dwarf candidates
were imaged with the NACO instrument of the VLT-UT4 (see instrument
description in Chauvin et al. 2010). Here we used the unique
  capability offered by NACO at the VLT to sense the wavefront in the
  near-infrared with the N90C10 dichroic ($90\%$ of the flux
  transmitted to the wavefront sensor and $10\%$ to CONICA). This mode
  is dedicated to the sharp imaging of very red sources
  $\rm{V}-\rm{K}\ge 6$ (M5 or later spectral type). All sources were
  imaged in $K_{s}$ bands (without saturation) in average seeing
  conditions of 0.8~$\!''$.  2MASS0652-57, which was resolved as a
  close binary, was also imaged in $J$ and $L~\!'$-bands.  The
  corresponding Strehl ratios, \textit{FWHM} (full
  width at half maximum of the diffraction limited point spread function) and other observing parameters in each band
  are given in the Table~2. The AO IR sensing allowed us to close the
  adaptive optics loop on all sources.  In Fig.~\ref{fig:image}, we
  have reported respectively all final images.
\begin{table*}[t]
\caption{Relative photometry and astrometry of 2MASS0652-57\,AB}
\begin{center}
\small
\begin{tabular}{llllllll}     
\noalign{\smallskip}\hline
\noalign{\smallskip}\hline  \noalign{\smallskip}
UT-Date            &  $\Delta J$     &  $\Delta K_s$      &  $\Delta L'$   & separation      & PA            & platescale    &  True North        \\
                   & (mag)           & (mag)             & (mag)         & (mas)           & (deg)         & (mas)         & (deg)              \\
\noalign{\smallskip}\hline  \noalign{\smallskip}
2006-01-08         &  $0.35\pm0.20$  & $0.34\pm0.10$     & $0.30\pm0.10$ & $228\pm6$       & $233.1\pm1.7$ & $27.02\pm0.06$ & $0.12\pm0.13$     \\
2010-08-21         &                 & $0.31\pm0.10$     &               & $222\pm7$       & $224.1\pm1.8$ & $27.02\pm0.05$ & $0.0\pm0.10$     \\
\noalign{\smallskip}\hline                  \noalign{\smallskip}
\end{tabular}
\end{center}
\end{table*}

Classical cosmetic reduction including bad pixels removal,
flat-fielding, sky substraction and shift-and-add, was made with the
\textit{Eclipse}\footnote{http://www.eso.org/projects/aot/eclipse/}
reduction software developed by Devillar (1997). Median filtering by a
kernel of $3 \times 3$ pixels was applied to correct for remaining hot
pixels. Finally, a high-pass filtering to remove spatial
  frequencies higher than $3\times\textit{FWHM}$ was applied to
search for fainter sources close to the target. The same reduction
strategy was used for all filters. For each epoch, the calibration of
the S27 mean platescale and the true north orientation was obtained
using the $\theta$ Ori C field observed with HST by McCaughrean \&
Stauffer (1994). The same set of stars (TCC058, 057, 054, 034 and 026)
were observed with the same observing set-up to avoid introducing any
systematic errors (see Table~5 of Chauvin et al. 2010, and Table~2 of
Chauvin et al. 2012).

\section{Results}

\subsection{Detection Limits}

\begin{figure}[t]
\centering
\vspace{-0.1cm}
\includegraphics[width = \columnwidth]{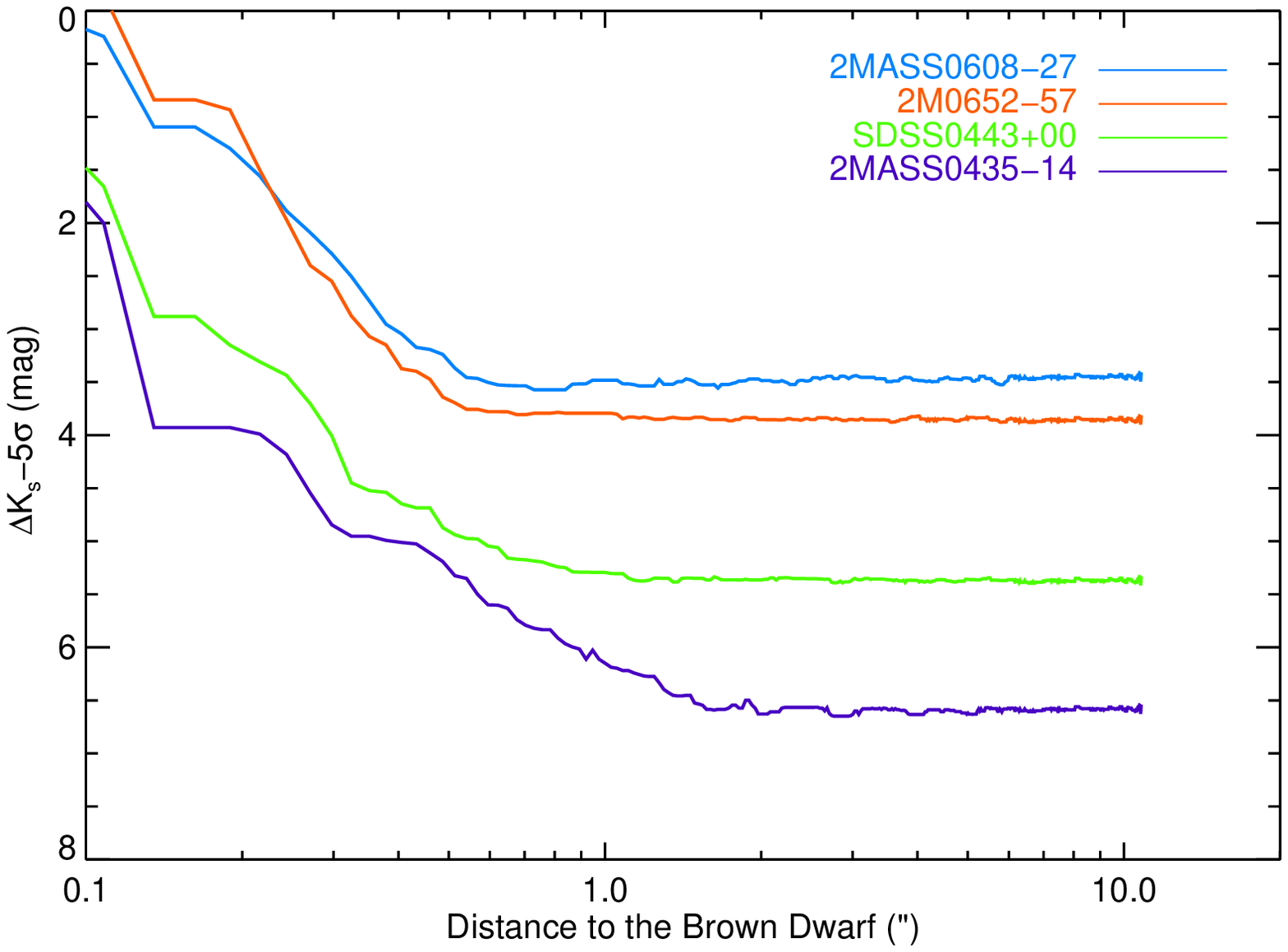}\hspace{0.5cm}
\includegraphics[width = \columnwidth]{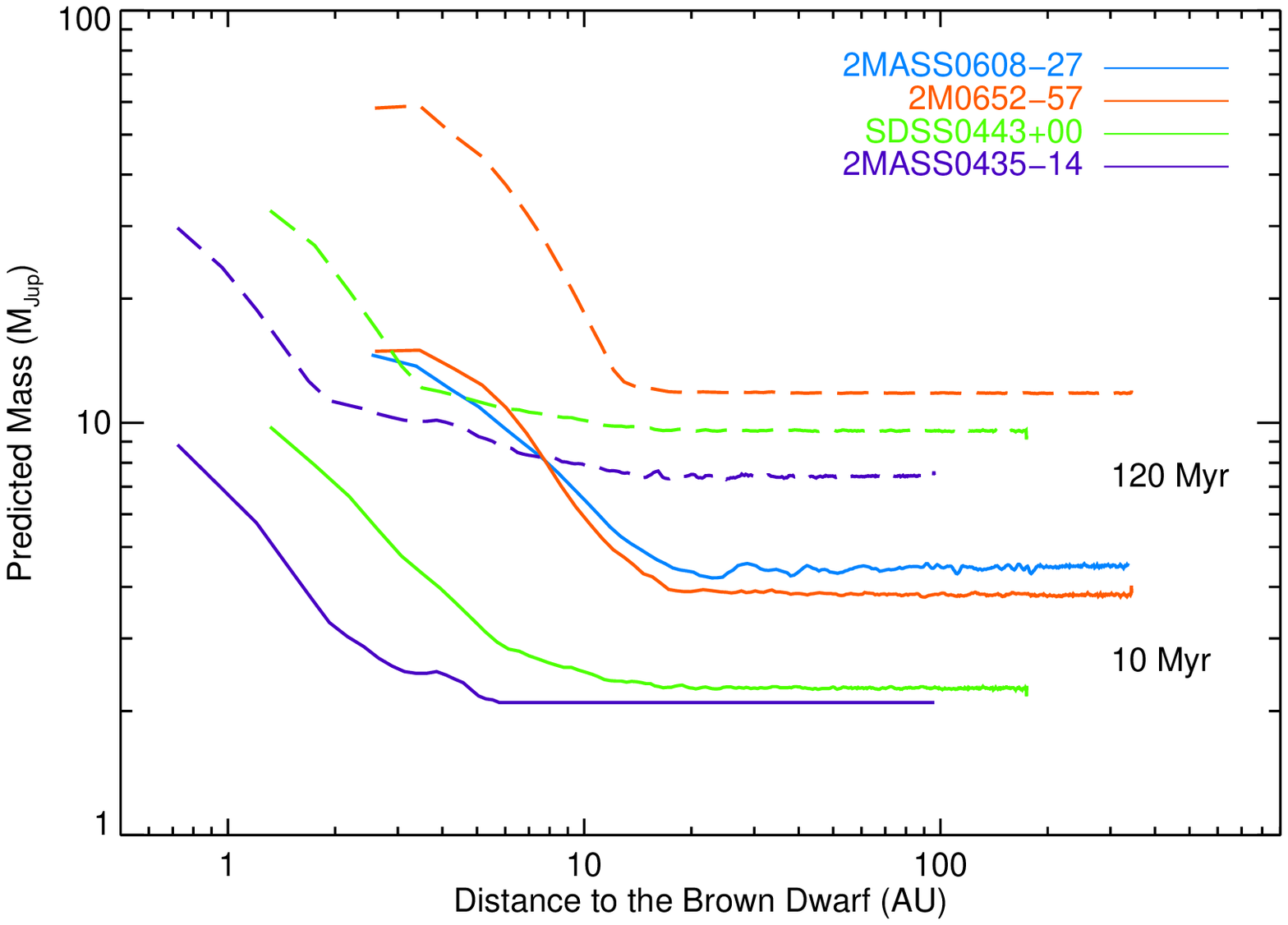}
\caption{\textit{Top:} VLT/NACO detection limits in $K_s$-band of
  2MASS0652-57, SDSS0443+00, 2MASS0608-27 and 2MASS0435-14, given here
  up to $10\,\!''$. \textit{Bottom:} Detection limits converted in
  terms of predicted masses as a function of the projected physical
  separations considering both the DUSTY theoretical model predictions
  for ages of 10 and 120~Myr and the target properties reported in
  Table~1. For 2MASS0608-27 probable member of the $\beta$ Pictoris
  group, only the age of 10~Myr is given. For 2MASS0652-57, part of
  the image with the binary companion was masked for the detection
  limit determination.}
\label{fig:detlim}
\end{figure}

For each brown dwarf candidate, the 5$\sigma$ detection limit was
estimated based on the noise calculated within azimuthal rings of
increasing radii centered on the source itself. The noise curve
was then divided by the primary star maximum flux. The results are
reported for the $K_s$-band filters in Fig.~\ref{fig:detlim},
(\textit{Top}). The detection performances vary as a function of both
the AO correction (related to the target near-IR brightness and the
atmospheric conditions) and the total observing time spent per target
to lower the background noise (see Table~2).
\begin{figure}[t]
\centering
\vspace{-0.1cm}
\includegraphics[width = \columnwidth]{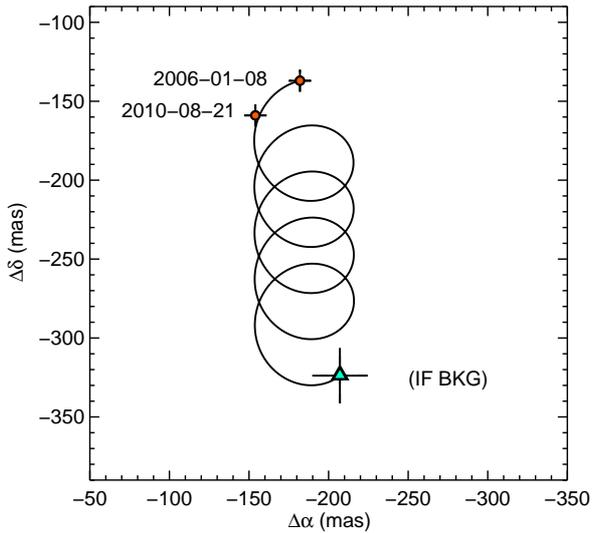}

\caption{VLT/NACO measurements with uncertainties in the offset
  positions of 2MASS0652-57\,B relative to A, obtained on January 8,
  2006 and August 21, 2010 (\textit{circle} data points). The \textit{solid line} gives the expected
  variation in the offset positions of B relative to A, if B is a
  background stationary object. The \textit{triangle} data point
  gives the predicted position for a background object on August 21,
  2010. Error bars take into account uncertainties on the initial
  offset position of B relative to A on January 8, 2006
  and on the proper and parallactic motions of 2MASS0652-57 reported by
  Faherty et al. (2012), see Table~1. }

\label{fig:astro}
\end{figure}

Owing to the challenge of observing faint near-IR targets with NACO,
the typical angular resolution achieved was about 95~mas, larger than
the diffration limit of the telescope in $K_s$ as the typical Strehl
correction was lower than 20\%.  The presence of close companions was
therefore probed between $0.1-10~\!''$ (i.e $2-250$~AU for a distance
of 20~pc). Based on the DUSTY theoretical models of Chabrier et
al. (2000), the detection limits were converted in terms of masses
considering both the target near-IR brightness and distance (reported
in Table~2) and the ages of 10 and 120~Myr, respectively (see
Fig.~\ref{fig:detlim}, \textit{Bottom}). They are sensitive down to
the planetary mass regime for these ages, but will be degraded to the
brown dwarf regime for older ages up to 500~Myr. Our best detection
limit reaches 2.5~M$_{\rm{Jup}}$ at 3~AU for 10~Myr. The young brown
dwarf candidate SDSS0443+00 was observed by Delorme et al. (2012) with
NACO in $L~\!'$ in December 2009 with deeper detection limits at
larger separation (see Fig.~6 of Delorme et al. 2012), but no faint
companions were resolved.

\subsection{2MASS0652-57, a young brown dwarf binary?}

Among the four young brown dwarf candidates, we discovered that
MASS0652-57 was a $\sim230$~mas visual pair. Both components were
resolved in $K_s$, but also in $J$ and $L'$-bands (see Fig.~1). The
relative position and photometry between both components were
estimated using a classical function fitting (gaussian and moffat) in
all filters for each individual components. The deconvolution
algorithm of V\'eran \& Rigaut (1998) was applied to check the impact
of the companion flux contamination. The reported uncertainties are
dominated by the dispersion of our results estimated from different
image set and related to the fluctuation of the AO correction. The
results are given in Table~3. The astrometric results are shown in
Fig.~3.

Considering the parallactic and proper motion of 2MASS0652-57 and
given the large time span between our observing epochs (4.5~yrs), we
can easily reject a stationary background contaminate. We used a
$\chi^2$ probability test of $2 \times N_{\rm{epochs}}$ degrees of
freedom following the approach described by Chauvin et
  al. (2010). We find a probability that the secondary component is a
background source lower than $10^{-5}$. We confirm that the system is
physically bound as the orbital curvature is unambiguously detected by
our two astrometric measurements. This result excludes an edge-on
orbital configuration.

Owing to the probability density distribution of the
  projected physical separation to semi-major axis ratio for a companion to a
  star (Brandeker et al. 2006), we estimate a probable semi-major axis
  of $7.3\pm3.6$~AU for 2MASS0652-57\,AB. DUSTY evolutionary
models predict a mass of 10-12, 15-20, 30-40 and 65-75~M$_{\rm{Jup}}$
for each component for ages of 10, 50, 120 and 500~Myr, respectively,
considering the system's properties reported in Table~1 and 2. It
would imply a $q \sim 0.7-0.8$ mass ratio and a period of
\textbf{$20-200$~yrs} depending on the system's age.  There is a
current lack of spectrocopic or kinematics constraints to firmly
confirm or reject the youth of this system. 2MASS0652-57 has a well
defined parallax and proper motion from Faherty et al 2012, however it
lacks a radial velocity measurement.  There is a low-resolution, low
SNR optical spectrum available from Cruz et al. (2007), but a
significant precise radial velocity can not be extracted from the
current data.  As such we can not conduct a full UVWXYZ analysis to
determine membership in nearby moving groups.  Given the proper motion
and distance measurement, there is a strong indication of kinematic
youth; however without a radial velocity we cannot constrain the age
in any significant way.  If confirmed as young and if the
  orbital period remains reasonably short for an astrometric
  monitoring, the
physical properties of the system could provide interesting
constraints on evolutionary model predictions over a decade or two
(Bonnefoy et al. 2009; Dupuy et al. 2009, 2010; Konopacky et
al. 2010).

\bibliographystyle{aa}

\begin{acknowledgements}

We thank Christine Ducourant, Rama Texeira and Joshua Schlieder for
the useful discussion about the status of these young brown dwarf
candidates and Alexis Brandeker for his meaningful comments as a referee for this research
note. We want to also thank the staff of ESO-VLT for their support at
the telescope. This publication has made use of the SIMBAD and VizieR
database operated at CDS, Strasbourg, France.  Finally, we acknowledge
support from the French National Research Agency (ANR) through project
grant ANR10-BLANC0504-01 and the {\sl Programmes Nationaux de
  Plan\'etologie et de Physique Stellaire} (PNP \&\ PNPS), in France.

\end{acknowledgements}


\end{document}